\documentclass[a4paper,10pt]{article}
\usepackage{enumerate}
\usepackage{color}
\usepackage[utf8]{inputenc} 
\usepackage[english]{babel}
\usepackage[T1]{fontenc}
\usepackage{graphicx}
\usepackage{amsfonts,amssymb,amsmath,latexsym,amsthm}
\usepackage{textcomp}
\usepackage[pdftex]{hyperref}
\usepackage{geometry}
\usepackage{color}
\DeclareMathOperator{\sech}{sech}

\title{Spring-mass behavior of solitons under the influence of an external force field within the modified Korteweg-de Vries equation}
\author{Marcelo V. Flamarion$^{1,*}$, Efim Pelinovsky$^{2,3}$ and Ioann Melnikov$^{2,3}$}
\date{}

\begin{document}
\maketitle
\begin{center}

\vspace{0.3cm}
{\footnotesize $^1$Departamento Ciencias—Secci{\' o}n Matem{\' a}ticas, Pontificia Universidad Cat{\' o}lica del Per{\' u}, Av. Universitaria 1801, San Miguel 15088, Lima, Peru \\
corresponding author*: mvellosoflamarionvasconcellos@pucp.edu.pe}

{\footnotesize $^{2}$Institute of Applied Physics, 46 Uljanov Str., Nizhny Novgorod 603155, Russia. \\
$^{3}$Faculty of Informatics, Mathematics and Computer Science, HSE University, Nizhny Novgorod 603155, Russia. }


\end{center}


\begin{abstract} 
We investigate the interaction of solitons with an external periodic field within the framework of the modified Korteweg-de Vries (mKdV) equation. In the case of small perturbation a simple dynamical system is used to describe the soliton behaviour. Equilibrium points of this dynamical system are computed when the external force travels at a constant speed. Assuming that the external force moves with sinusoidal speed, we demonstrate that the soliton behavior is qualitatively similar to the constant-speed case.  Besides, a resonant frequency is derived from the asymptotic theory without using the classical broad force approximation.  The results obtained from the dynamical system are compared with fully direct numerical simulations, which reveal that the soliton solution exhibits spiral-like behavior in the soliton amplitude versus soliton phase space. Moreover, when the external force oscillates at the resonant frequency, the trajectories in the soliton phase versus soliton amplitude exhibit chaotic behavior.

	\end{abstract}

\section{Introduction}
The interaction of solitary waves with external wave fields has been extensively studied in the literature using various equations, including the Korteweg–de Vries (KdV) equation \cite{Akylas:1984, Wu:1982, Wu:1987, Wu1, Milewski:2004, Grimshaw:1986, Grimshaw:2016, Grimshaw:2019, Choi:2008, Ermakov:2019}, the Benjamin-Ono equation \cite{Flamarion-Pelinovsky:2023, Flamarion:2024, Matsuno:1993a, Matsuno:1993b}, the modified KdV (mKdV) equation \cite{Chaos:FP}, the Gardner equation \cite{Grimshaw:2002, Flamarion-Pelinovsky:2022a}, and non-integrable models  \cite{Flamarion:2022b, Chowdhury :2018, Flamarion:2023a}. Understanding the response of wave fields to external forces is important, as these forces can represent a direct pressure distribution on the wave field, variable topography, or even random noise \cite{Akylas:1984, Baines:1995}. From a physical perspective, forced equations are utilized as models to investigate a variety of nonlinear wave phenomena. These include ship wakes, fluid flows around obstacles, waves on the surface of a conductive fluid under the influence of an electric field \cite{Perelman:1974}, quantum-dimensional film waves, and elastic waves in solids \cite{Pavlov:1998}. Additionally, they describe trapped waves \cite{Malomed:1993, Grimshaw:1993, Grimshaw:1994, Lee:2018, Kim:2018, LeeWhang:2018}, as well as wave turbulence in free-surface magnetohydrodynamic and electro-hydrodynamic systems \cite{Kochurin:2020a, Kochurin:2019, Kochurin:2025, Kochurin:2025a, Dmitriev:2023, Kochurin:2022b}.  

Within the KdV framework, a comprehensive asymptotic and numerical analysis of the forced KdV equation was first conducted by Grimshaw et al. \cite{Grimshaw:1996}, who demonstrated strong agreement between theoretical predictions and numerical results and established conditions for wave trapping.  Most studies have concentrated on localized external forces, with only a few exploring the effects of periodic external forces \cite{Malomed:1993, Flamarion-Pelinovsky:2022a}. The introduction of a periodic force poses a substantial mathematical challenge, as it produces radiation throughout the domain as the force moves. This pervasive radiation complicates the task of distinguishing between radiation arising from the interaction of the solitary wave with the external force and that originating solely from the periodic force. The pioneering work on solitary wave interactions with a periodic external force was conducted by Malomed et al. \cite{Malomed:1993}, who introduced a change of variables to filter out the linear wave interactions generated by the periodic force. Building on this approach, Flamarion and Pelinovsky \cite{Flamarion-Pelinovsky:2022a} later applied the method to the Gardner equation, establishing conditions under which waves could be trapped by the external force.

This work aims to investigate, both numerically and asymptotically, the interaction between solitons and an external force with variable speed, using the forced modified Korteweg–de Vries (mKdV) equation with positive cubic nonlinearity. The mKdV equation supports breathers—traveling oscillatory wave packets—as well as solitons of both polarities \cite{Pelinovsky:2002, Pelinovsky:2020}. In this work, we qualitatively and quantitatively compare asymptotic and numerical results and discuss conditions for soliton trapping.  Specifically, we demonstrate that while the asymptotic theory predicts the existence of perfect resonance between the soliton and the external term, with closed orbits in the soliton-amplitude vs. soliton-phase space, this behavior is not observed in direct numerical simulations. Instead, numerical simulations reveal that the soliton behaves as an unstable spiral, spiraling outward more rapidly as the external force wavelength increases. Notably, when the wavelength of the periodic force is short, the soliton nearly exhibits perfectly closed trajectories. Besides, when the external force oscillates with time,  a resonant frequency is derived.

This article is organized as follows. In section 2 we present the mathematical formulation of the problem. Asymptotic results  are presented in section 3 and  numerical results in section 4. The conclusion is presented in section 5.
%
%
%

\section{The mathematical model}
We consider the mKdV equation in the presence of a weak force that travels at variable speed ($v(t)$) 
\begin{equation}\label{mfKdV1}
U_{t} +cU_x+6U^{2} U_x+U_{xxx}=\epsilon f_{x}\Big(x-\int v(t) dt\Big),
\end{equation}
to investigate the interaction of solitons with a external forcing field. Here, we denote by $U(x,t)$  the wave field, $f(x)$  the external forcing that travels with variable speed $v(t)$, $c$ the long-wave propagation speed and $\epsilon$  a small positive parameter.  It is convenient to rewrite equation (\ref{mfKdV1}) in the reference frame moving with the external force. This can be accomplished by considering the change of variables:
\begin{equation}
x' = x - \int v(t) \, dt, \quad t' = t.
\end{equation}
With this substitution, the equation is transformed to a frame that adapts to the motion of the external force, simplifying further analysis. In the new coordinate system, dropping the primes, equation (\ref{mfKdV1}) reads
\begin{equation}\label{mfKdV}
U_{t} +\Delta(t)U_x+ 6U^{2} U_x+U_{xxx}=\epsilon f_{x}(x),
\end{equation}
where 
\begin{equation} \label{deviation}
\Delta(t)=c-v(t),
\end{equation}
is the deviation speed. This important parameter measures the difference between the linear long-wave speed and the speed of the external forcing \cite{Grimshaw:1996}.
Equation (\ref{mfKdV}) conserves the total mass $(M(t))$, with
\begin{equation}\label{mass}
\frac{dM}{dt}=0, \mbox {where } M(t)=\int_{-\infty}^{\infty}U(x,t)dx.
\end{equation} 
Additionally, the rate of change of momentum $(P(t))$ is balanced by the external forcing through the equation
\begin{equation}\label{momentum}
\frac{dP}{dt}=\int_{-\infty}^{\infty}U(x,t)\frac{df(x)}{dx}dx, \mbox {where } P(t)=\frac{1}{2}\int_{-\infty}^{\infty}U^{2}(x,t)dx.
\end{equation} 
In the absence of an external forcing, the mKdV admits  solitons as solutions   which are given by the expressions \cite{Pelinovsky:2002},
\begin{equation}\label{solitary}
U(x,t)=a\sech(a\mathbf{\Phi}), \mbox{  where }  \;\ \mathbf{\Phi}=x-qt-x_0, \;\ q = c-v+a^2,
\end{equation}
where $a$ is amplitude of the soliton,  $x_0$ is its the initial position of the crest. Here, $a$ is allowed to be negative, which represents depression solitary waves.

Flamarion and Pelinovsky \cite{Chaos:FP} demonstrated that, under the influence of a weak external force, the soliton undergoes an adiabatic transformation. They further showed that the soliton parameters, specifically its amplitude and phase, evolve according to the corresponding dynamical system
\begin{align} \label{DS0}
\begin{split}
& \frac{{da}}{dT} = a\int_{-\infty}^{\infty}\sech(a\mathbf{\Phi})\frac{df}{d\mathbf{\Phi}}(\mathbf{\Phi}+\mathbf{\Psi})d\mathbf{\Phi}, \\
& \frac{d{\mathbf{\Psi}}}{dT}=\Delta(T)+a^2, 
\end{split}
\end{align}
where 
\begin{equation}\label{solitary}
U(\mathbf{\Phi},T)=a(T)\sech(a(T)\mathbf{\Phi}), \mbox{  where }  \;\ T=\epsilon t, \;\ \mathbf{\Phi}=x-\mathbf{\Psi}(T), \mbox{ and } \mathbf{\Psi}(T) =x_0+\frac{1}{\epsilon}\int q(T)dT.
\end{equation}

In fact, system (\ref{DS0}) is a truncated version that excludes small second-order terms. Following the approach of similar studies \cite{Grimshaw:1994, Grimshaw:1996, Grimshaw:1993, Grimshaw:2002}, we begin by analyzing this simplified system, as it offers a more tractable framework for analytical investigation. 

In this analysis, we incorporate the periodic external force within the scope of asymptotic theory. Notably, when the external force has a broader spatial scale than the soliton length, the asymptotic theory applies to any function  $f(x)$, beyond just those that vanish at infinity. This framework also extends to periodic external forces with relatively small wavenumbers. Consequently, we define the external force as
\begin{equation}\label{periodic}
f(x)=A\sin(qx),
\end{equation}
where $A$ is the external force amplitude and $2\pi/q$ is its wavelength.  Notice that if $U(x,t)$ is a solution to equation (\ref{mfKdV}) with an external force $f(x)$, then $-U(x,t)$ also satisfies the equation when this term is replaced by $-f(x)$. Consequently, we focus our analysis on the case where $A > 0$, as similar results hold for $A < 0$. This approach allows us to concentrate on positive amplitude scenarios, recognizing that a sign change in $A$ simply results in a mirrored solution within the solution space.

\section{Asymptotic results}

\subsection{Soliton interactions with a periodic external force with constant speed}
\begin{figure}[h!]
	\centering	
	\includegraphics[scale =0.49]{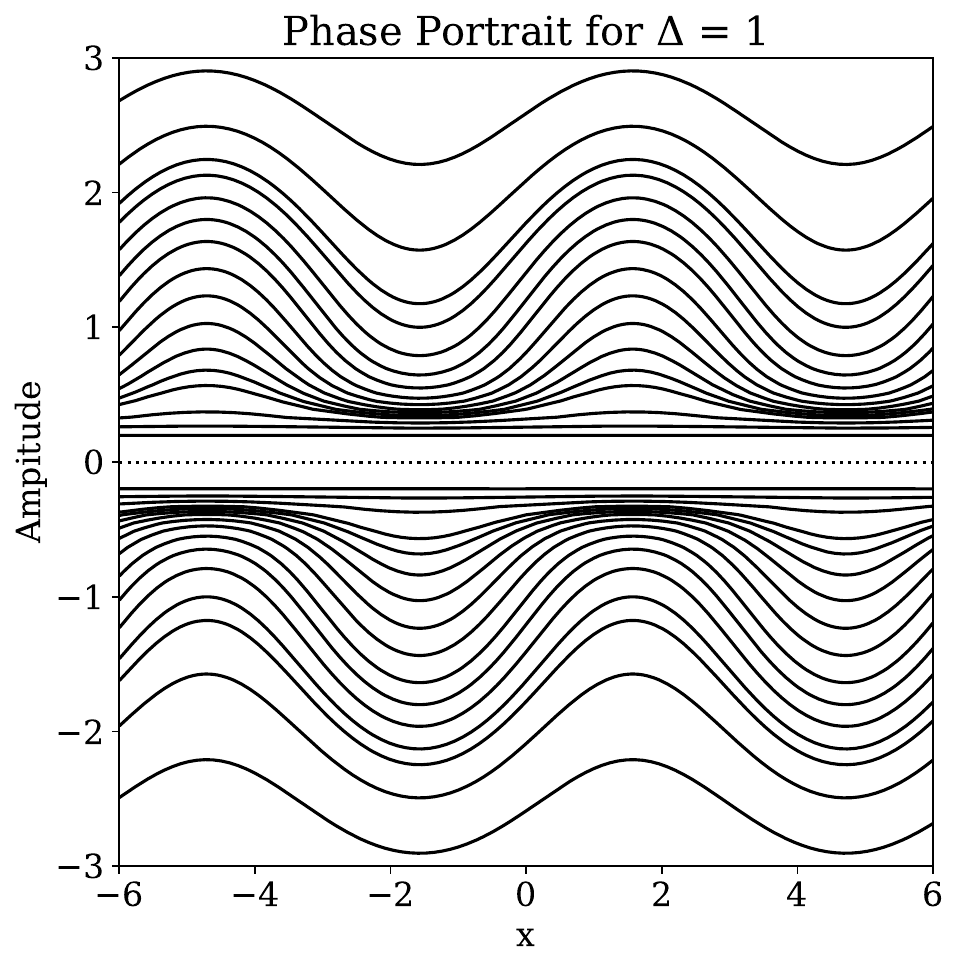}
 	\includegraphics[scale =0.49]{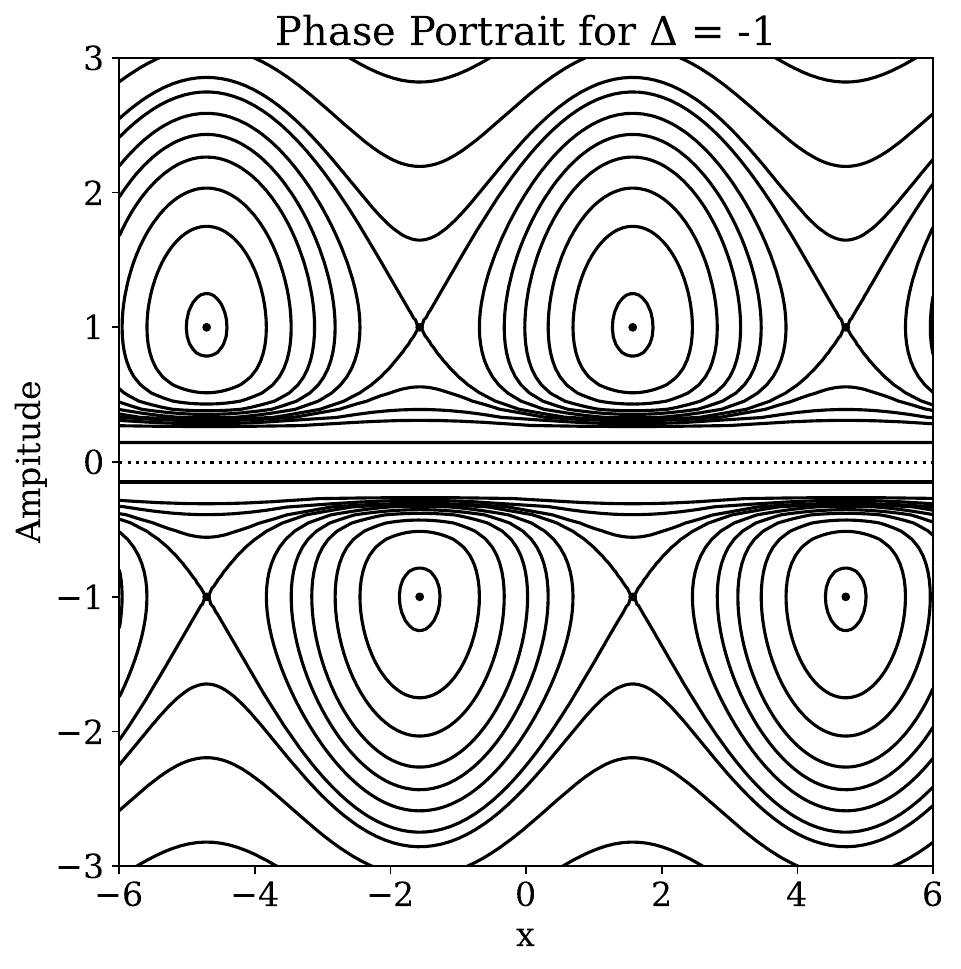}
	\caption{Phase portraits for the dynamical system (\ref{DS_without_broad_force}). Circles correspond to centres and  crosses to saddles ($A=\Delta=q=1$).}
	\label{Fig1}
\end{figure}

This analysis parallels the findings for the Gardner equation, as reported by Pelinovsky and Flamarion \cite{Flamarion-Pelinovsky:2022a}. We begin by assuming that the external term moves with a constant velocity. {{Previously, the literature primarily addressed cases of limiting interactions between solitons and external forces, where the length of one significantly exceeded that of the other. In this study, we relax this assumption by investigating the general interaction scenario with a given sinusoidal force (\ref{periodic}). Notice that in our case, the integral in first equation of (\ref{DS0}) can be computed using the residue technique, see \cite{Granshtein} (section 3.981). So system (\ref{DS0}) becomes }} (for simplicity we replace $\Psi$ on $x$)
\begin{align} \label{DS_without_broad_force}
\begin{split}
& \frac{da}{dT}=Aq\pi \cos(qx) \sech \Big(\frac{q \pi}{2a}\Big), \\
& \frac{dx}{dT} = \Delta + a^2.
\end{split}
\end{align}
The behavior of system (\ref{DS_without_broad_force}) for a positive value of $\Delta$—corresponding to the case where the external disturbance travels faster than the long-wave speed ($v > c$)—is relatively straightforward. In this scenario, the soliton behaves as a flyby soliton for any amplitude and position. The corresponding phase plane is presented on the left side of Fig. \ref{Fig1}, and this case will not be discussed further.

The system (\ref{DS_without_broad_force}) undergoes a bifurcation when the sign of the $\Delta$ parameter changes. In this case, the dynamical system (\ref{DS_without_broad_force}) exhibits an infinite number of equilibrium points corresponding to the local extrema of the function $f(x)$. These equilibrium states alternate between saddles and centers, which we refer to as resonant solitons. For a given negative value of $\Delta$, the amplitude $a_0$ of the resonant soliton and the position $x_0$ of its crest are independent of the amplitude $A$ and are given by the following expressions
\begin{equation} \label{equilibrium}
a_{0}^{\pm}=\pm\sqrt{-\Delta}, \mbox{ and } x_{0}=(2n+1)\frac{\pi}{2q}, \mbox{ $n\in \mathbb{Z}$}.
\end{equation}
The equilibrium point is a center if the disturbance and the soliton share the same polarity, while it is a saddle if their polarities are opposite, as illustrated in Fig. \ref{Fig1} (right panel).

At this stage, it is crucial to establish a connection between the solutions of the dynamical system (\ref{DS_without_broad_force}) and those of the modified Korteweg–de Vries (mKdV) equation (\ref{mfKdV}). In this context, a center corresponds to a soliton that is in exact resonance with the external term, remaining stationary and effectively locked in position. This equilibrium represents a captured solution where the soliton is synchronized with the external influence. The presence of closed orbits indicates that the soliton solutions oscillate around the external forcing term, following a repetitive trajectory. In such cases, the position of the soliton crest becomes a periodic function of time, reflecting the cyclical interaction with the external term. These closed orbits suggest that while the soliton experiences periodic motion, it remains bound within a certain region influenced by the external term. In contrast, saddle points represent solitons that are repelled by the external term. For these solutions, small perturbations cause the soliton to drift away from the external influence over time, indicating instability in the soliton position. This repulsive behavior highlights the critical role of the saddle points in delineating regions of stable and unstable interactions within the solution space.
Assuming that the external term moves at a constant speed, the dynamical system (\ref{DS_without_broad_force}) is autonomous. Moreover, system (\ref{DS_without_broad_force}) is conservative, possessing the first integral $F$.
\begin{equation}
F(x, a) = \int_{a_0} ^a (\xi^2 + \Delta)\cosh \Big ( \frac{q \pi}{2 \xi}\Big )d \xi - A \pi \sin(qx).
\end{equation}

It becomes much easier to work with the system under the assumption that the soliton width is narrow compared to the external force. In this case, system (\ref{DS0}) takes the Hamiltonian form
\begin{align} \label{DS}
\begin{split}
& \frac{da}{dT}=\pi\frac{df(x)}{dx}, \\
& \frac{dx}{dT} = \Delta(T)+a^2.
\end{split}
\end{align}  
Consequently, its solutions can be represented by a conserved Hamiltonian function, which provides a convenient framework to analyze the system behavior. The Hamiltonian $H(x,a)$, which is expressed as
\begin{equation}\label{streamfunction}
H(x,a) = -\pi f(x)+\Delta a+\frac{a^3}{3}.
\end{equation}

A key advantage of system (\ref{DS}) is that, through appropriate variable rescaling, most parameters can be eliminated, reducing the system to a single governing parameter. Specifically, we rescale the spatial coordinate $ x $  by the inverse of the wavenumber $q$, and the amplitude $ a $ by the factor $ (A\pi)^{1/3} $, where $ A > 0 $ represents the amplitude of the external force. This rescaling not only simplifies the expressions but also introduces a dimensionless parameter 
\begin{equation}
\widetilde{\Delta}= \frac{\Delta}{(\pi A)^{2/3}}.
\end{equation}
\begin{figure}[h!]
	\centering	
	\includegraphics[scale =1.05]{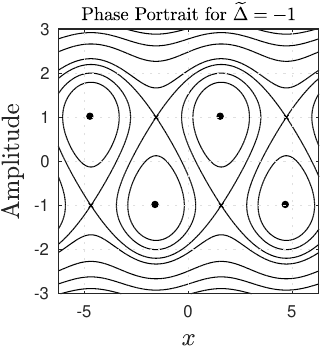}
	\includegraphics[scale =1.05]{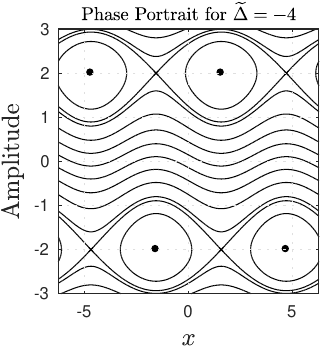}
	\caption{Phase portraits for the dynamical system (\ref{DS}). Circles correspond to centres and  crosses to saddles.}
	\label{Fig1b}
\end{figure}
In this new coordinates system, the  streamfunction reads
\begin{equation}\label{streamfunction}
H(x,a) = -\sin x+\widetilde{\Delta} a+\frac{a^3}{3}.
\end{equation}

Typical phase portraits of the dynamical system are displayed in Figure \ref{Fig1b} (for positive values of $ \widetilde{\Delta} $, there are no equilibrium points for (\ref{DS}) and we do not discuss it).  Additionally, for negative values of $ \widetilde{\Delta} $, the system predicts that the soliton solution may change its polarity, as trajectories cross the zero-amplitude level. This is the main topological difference between system (\ref{DS}) and system (\ref{DS_without_broad_force}). However, in the vicinity of the zero amplitude, system (\ref{DS}) and system (\ref{DS_without_broad_force}) must be treated carefully, since all terms in the forced mKdV equation become comparable in magnitude, leading to a breakdown of perturbation theory. Thus, the phase portrait should be interpreted with caution and primarily applied to cases where the soliton amplitude is large in modulus.


\subsubsection{Soliton interactions with an external forcing with variable speed}
In this section, we examine the case where the external term moves with a variable speed. Specifically, we  focus on the case when the external force varies according to a sinusoidal law. This scenario is particularly relevant to the problem of a periodic force acting on a nonlinear oscillator. The outcome is intuitive: under resonant conditions (due to the variable component of the velocity), oscillations can grow over time, ultimately causing the soliton to escape the trapping zone. Therefore, the oscillatory component of the external field motion disrupts the physical resonance between the external field and the soliton, weakening the interaction. We represent the variable speed as
\begin{equation}\label{sinus}
v(T) = c - V_{0} - V\sin(\omega T),
\end{equation}
where $V$ is the amplitude of the oscillation and $\omega$ the frequency. With this choice we have that 
\begin{equation}\label{dsinus}
\Delta(T) = V_{0} + V\sin(\omega T).
\end{equation}

\begin{figure}[h!]
	\centering	
	\includegraphics[scale =0.3]{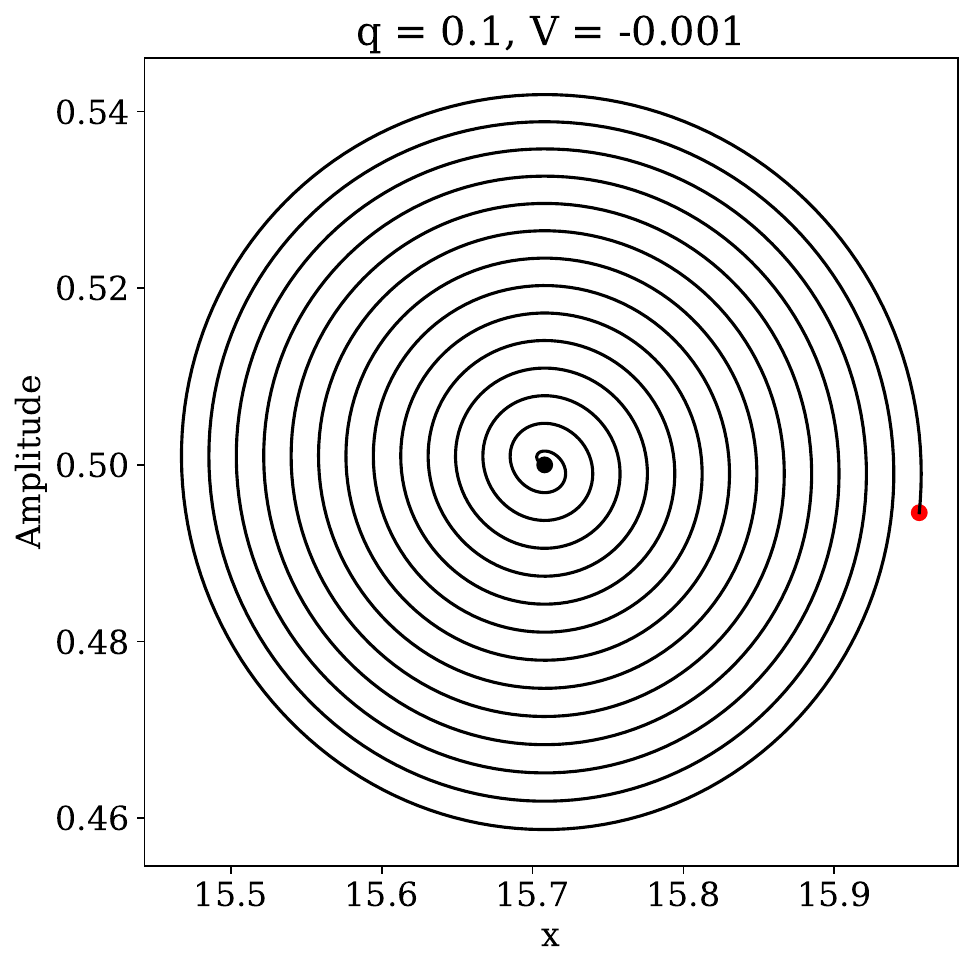}
 	\includegraphics[scale =0.3]{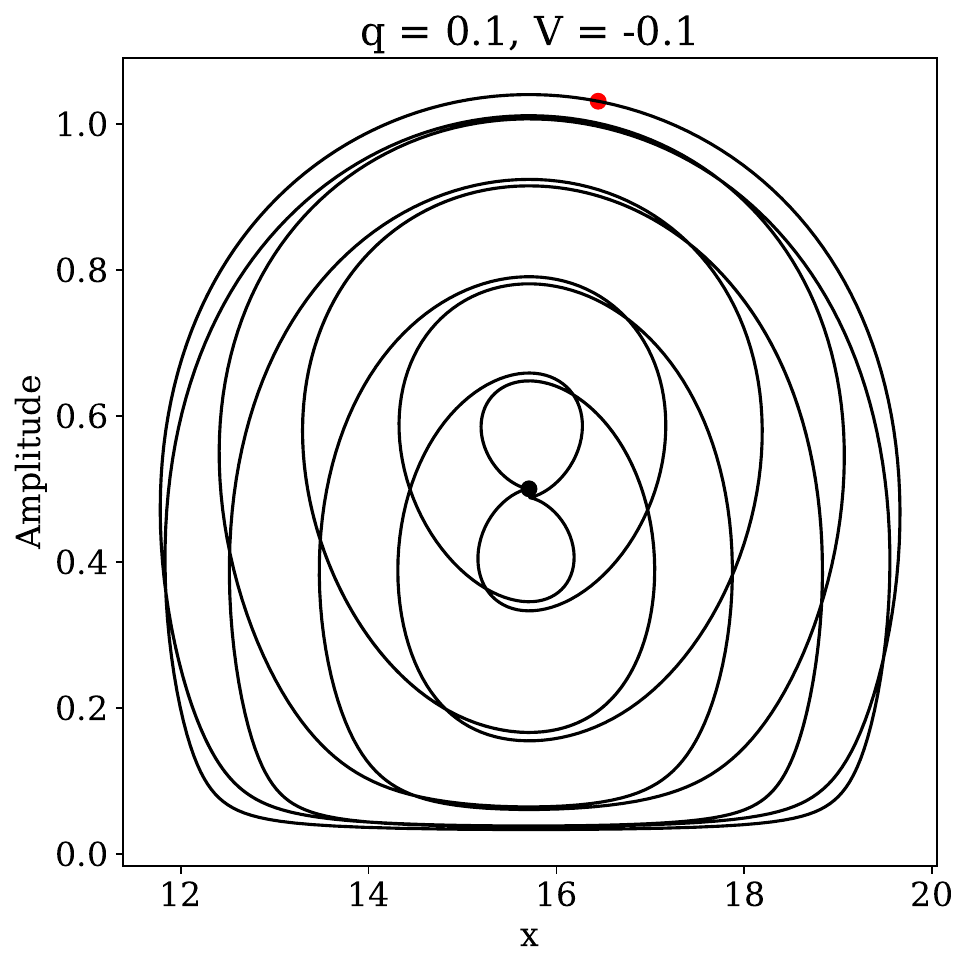}
	\includegraphics[scale =0.3]{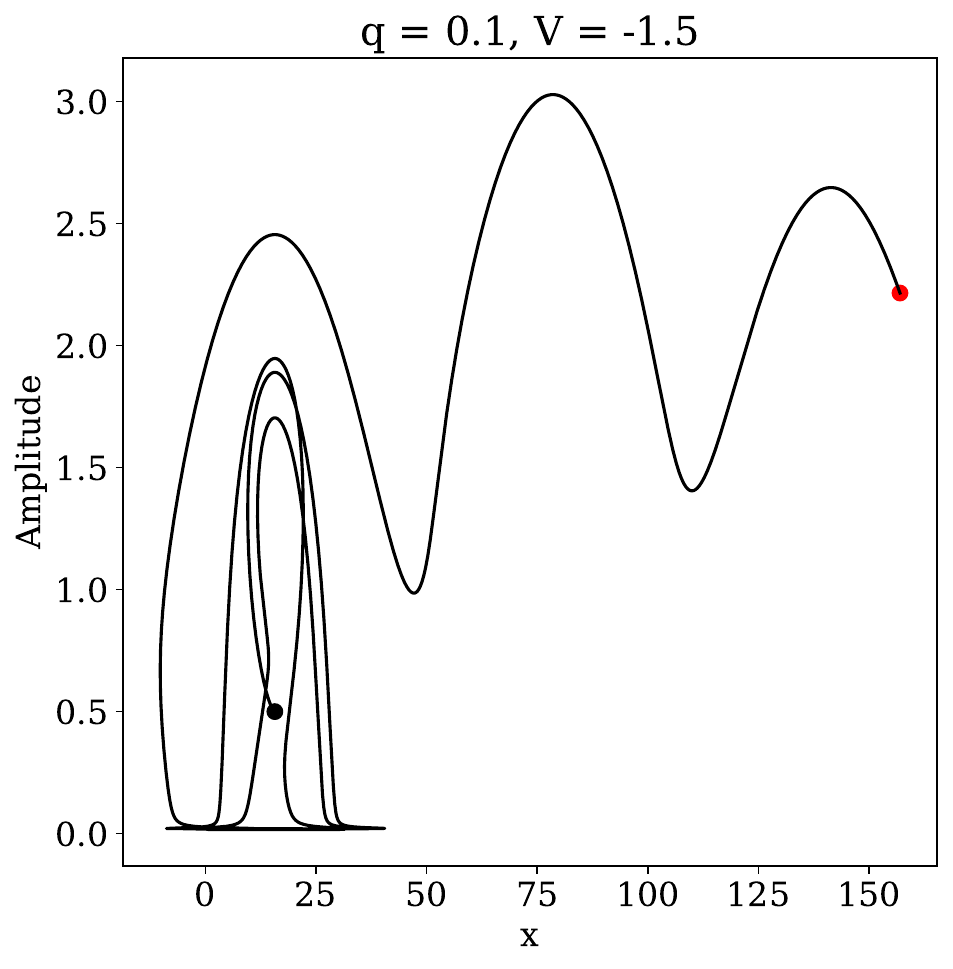}
	\caption{The system (\ref{DS_without_broad_force}) trajectory coming out of the black point (center for an unperturbed system) with parameters $V_0=-a_0^2=-0.25$, $A=1$, $\omega=\omega_0$. The red dot is the end point of the trajectory ($T\in[0,500]$). The change in variable velocity $V$ increases from left to right.}
	\label{Fig01}
\end{figure}

\begin{figure}[h!]
	\centering	
	\includegraphics[scale =0.3]{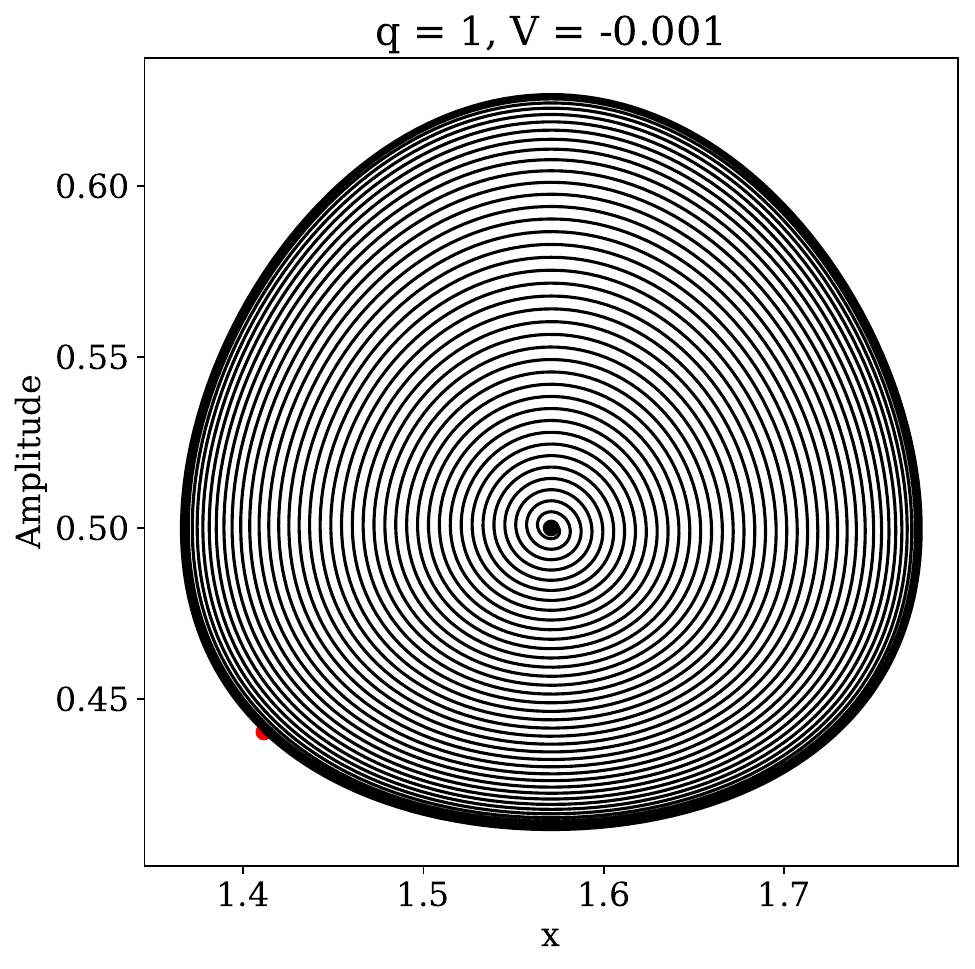}
	\includegraphics[scale =0.3]{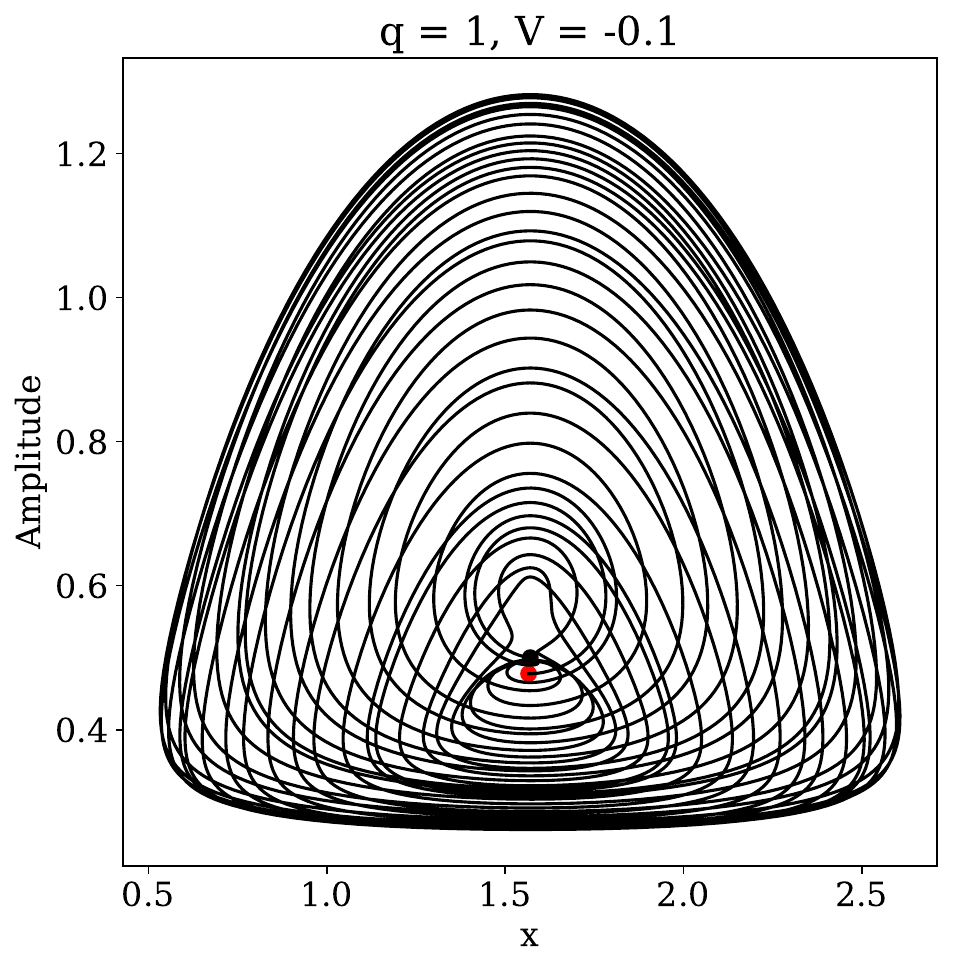}
        \includegraphics[scale =0.3]{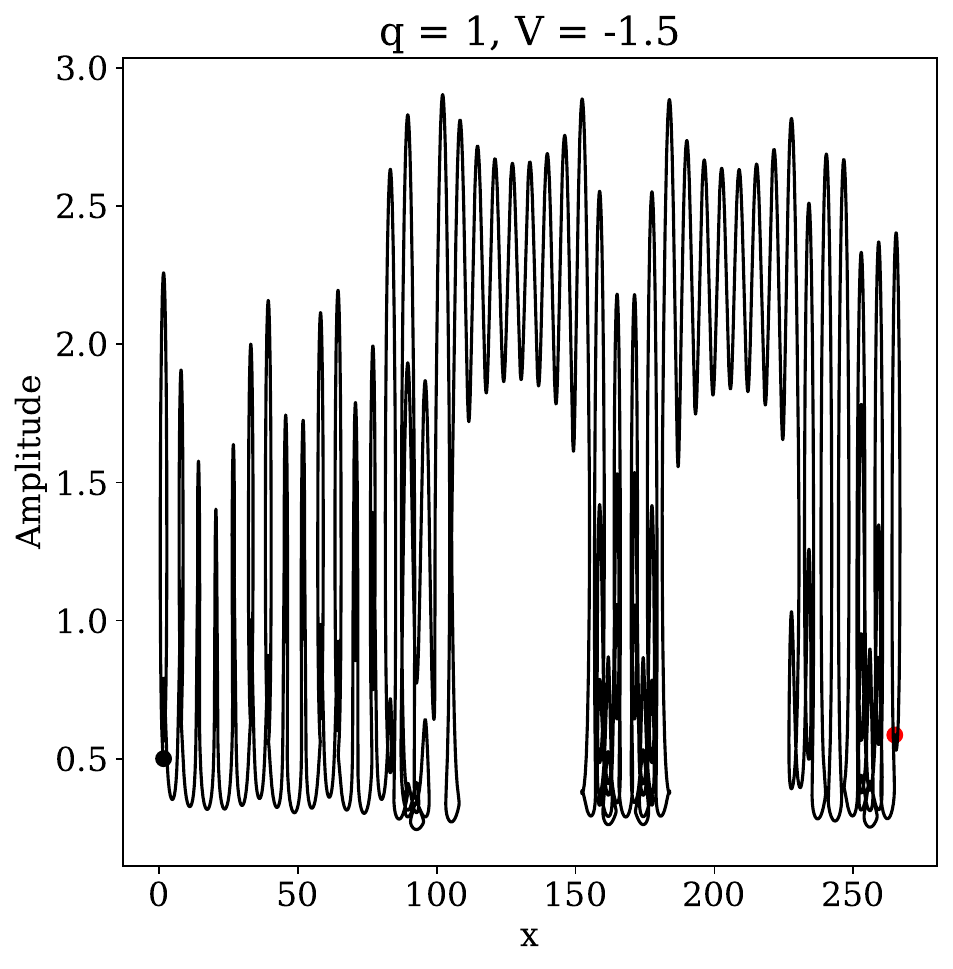}
	\caption{It is the same as in Figure \ref{Fig01}, but $q=1$.}
	\label{Fig02}
\end{figure}

Let us determine the value of the speed change frequency $\omega$ required for the system to exit the captured region. In this case, the linearized form of system (\ref{DS_without_broad_force}) near the center $(x_0, a_0)$ is given by:
\begin{equation} \label{linear}
\begin{split}
& \frac{d\tilde{x}}{dT} = 2\sqrt{-V_0}\tilde{a} + V\sin(\omega T), \\
& \frac{d\tilde{a}}{dT}=-Aq^2 \pi \sech \Big( \frac{k \pi}{2 \sqrt{-V_0}}\Big)\tilde{x},
\end{split}  
\end{equation}
where
\begin{equation}
    \tilde{x} = x - x_0,~~\tilde{a} = a - a_0.
\end{equation}
System (\ref{linear}) is trivially reduced to the equation of the classical oscillator
\begin{equation}
\frac{d^2 \tilde{a}}{dT^2} + \omega_{0}^2 \tilde{a} = f_0 \sin(\omega T),
\end{equation}
where
\begin{equation} \label{resonance}
\omega_0 = \sqrt{2\pi Aq^2 \sqrt{-V_0} \sech \Big( \frac{q \pi}{2 \sqrt{-V_0}} \Big )} \mbox{ and }  f_0 = V \pi Aq^2 \sech \Big( \frac{q \pi}{2 \sqrt{-V_0}}\Big ).
\end{equation}
With a small change in the external force velocity at the resonant frequency, we observe how the amplitude and coordinates begin to increase relative to the undisturbed center. However, due to the nonlinearity of the system, this does not persist over time (Fig. \ref{Fig01} and \ref{Fig02}, on the left). Of course, due to nonlinearity, it is not possible to achieve linear growth, because if we look at the first equation of system (\ref{DS_without_broad_force}), we have
\begin{equation} \label{ineq}
\Big |\frac{da}{dT}\Big| < Aq\pi, \text{ then } |a| < |a_0| + Aq\pi T~~\forall T >0,
\end{equation}
where $a_0$ is the initial amplitude. For the simplified system (\ref{DS}), inequality (\ref{ineq}) is not strict, and we can even explicitly find a perturbation (which, however, it  is not harmonic) that causes a linear increase in amplitude. For example, to obtain the solution $(x, a) = (0, Aq\pi T)$, the speed must change as $\Delta = -(Aq\pi T)^2$.

With an increase in the harmonic ($\omega = \omega_0$) change in velocity, the solution exhibits a quasi-periodic and trapped character (Figures \ref{Fig01} and \ref{Fig02}, center panels). For an even larger (in magnitude) value of $V$, the soliton can exit the trapped state and begin wandering along the phase plane (Figures \ref{Fig01} and \ref{Fig02}, right panels). It is also worth noting that for very small values of $V$, achieving a significant amplitude gain requires an excessively long time, which lies beyond the scope of the asymptotic theory.

\begin{figure}[h!]
	\centering	
	\includegraphics[scale =0.9]{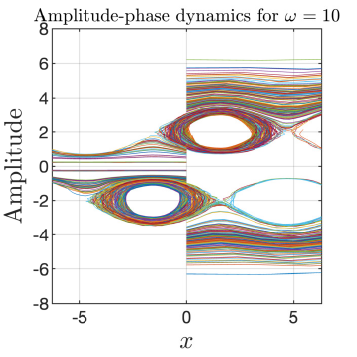}
       \includegraphics[scale =0.9]{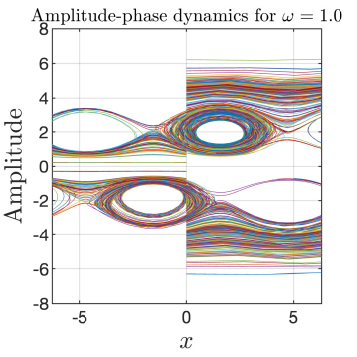}
	\includegraphics[scale =0.9]{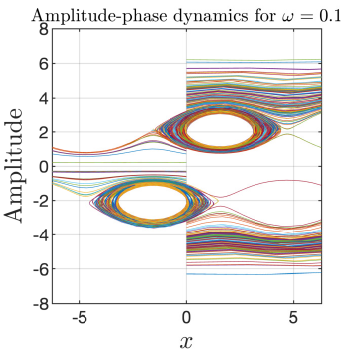}
	\caption{Amplitude-phase dynamics for different initial values of (\ref{DS_without_broad_force}) and variable speed (\ref{sinus}). Parameters: $V_0=-4$, $V=1$. Circles correspond to the starting point and squares to the ending points. }
	\label{Fig2}
\end{figure}

Notice that when the amplitude of the oscillator is close to zero ($ V \approx 0 $), the system of ordinary differential equation becomes nearly autonomous. In this regime, the influence of external forces or time-dependent effects is minimal, suggesting that the amplitude-phase dynamics should resemble those of the phase portraits shown in Figure \ref{Fig1} for the constant speed case. Consequently, the system behavior is expected to exhibit similar qualitative features, such as periodic or quasi-periodic motion, as observed in the simpler constant speed scenario (Figure \ref{Fig2}). 

\begin{figure}[h!]
	\centering	
	\includegraphics[scale =0.9]{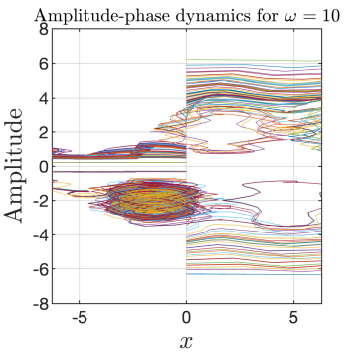}
 	\includegraphics[scale =0.9]{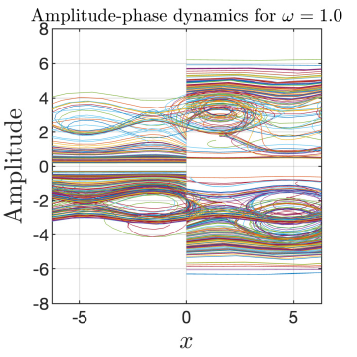}
	\includegraphics[scale =0.9]{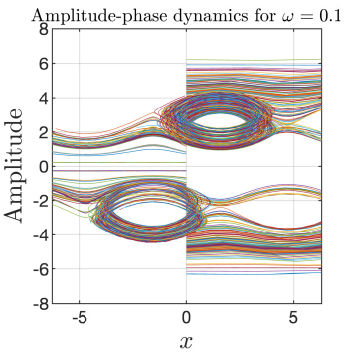}
	\caption{Amplitude-phase dynamics for different initial values of (\ref{DS}) and variable speed (\ref{sinus}). Parameters: $V_0=-4$, $V=8$. Circles correspond to the starting point and squares to the ending points.}
	\label{Fig7}
\end{figure}

Observe that as the frequency approaches zero ($ \omega \approx 0 $), the system of differential equations becomes nearly autonomous. In this low-frequency limit, time-dependent influences diminish, suggesting that the amplitude-phase behavior is to align with the characteristics of the phase portraits depicted in Figure \ref{Fig7} for the constant speed scenario. Consequently, the system is expected to exhibit analogous dynamical patterns, including steady or nearly steady trajectories, similar to those seen in the constant speed case.

\begin{figure}[h!]
	\centering	
	\includegraphics[scale =0.3]{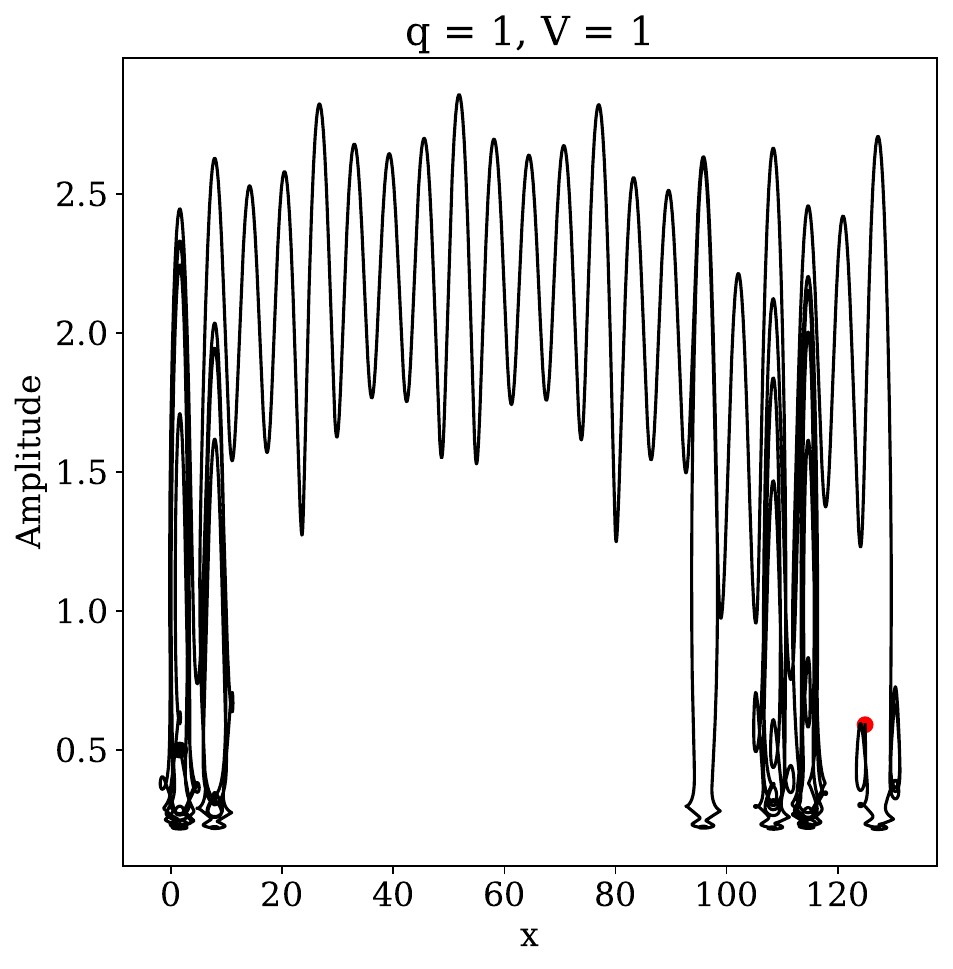}
 	\includegraphics[scale =0.3]{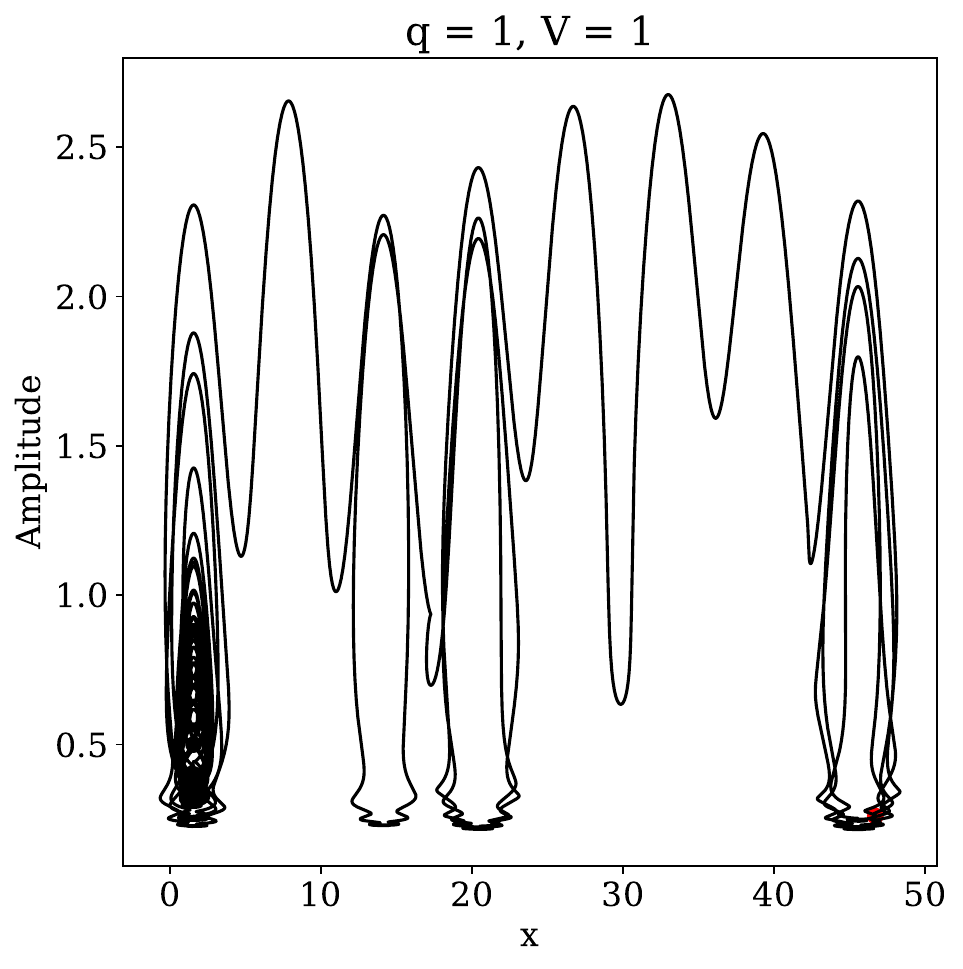}
	\includegraphics[scale =0.3]{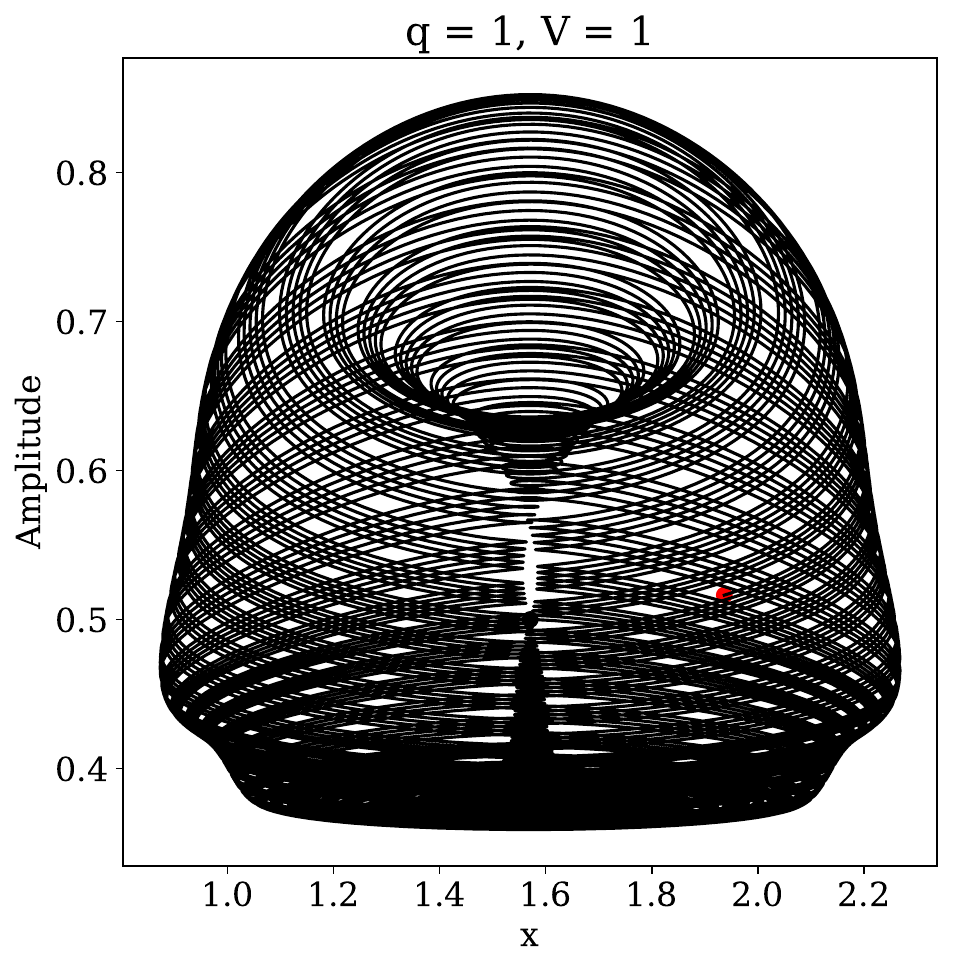}
	\caption{It is the same as in Figure \ref{Fig01}, but frequency from left to the right $\omega=1, 2$ and 3}.
	\label{Fig03}
\end{figure}

For arbitrary sets of parameters $\omega$ and $V$, the soliton motion resembles either a chaotic walk with varying amplitude (Fig. \ref{Fig03}) or trapped movement near the equilibrium state (Fig. \ref{Fig03}, right panel). Providing a rigorous mathematical justification for chaos in a perturbed conservative system is challenging. For instance, Lyapunov exponents, commonly used to characterize attractors through phase volume compression, are not directly applicable here, as the phase volume remains constant in a conservative system. However, given that system (\ref{DS_without_broad_force}), when considered with the following approximations, is no longer strictly conservative, it is reasonable to justify the wandering trajectories observed in the figures through further research based on these approximations.

\begin{figure}[h!]
	\centering	
	\includegraphics[scale =0.3]{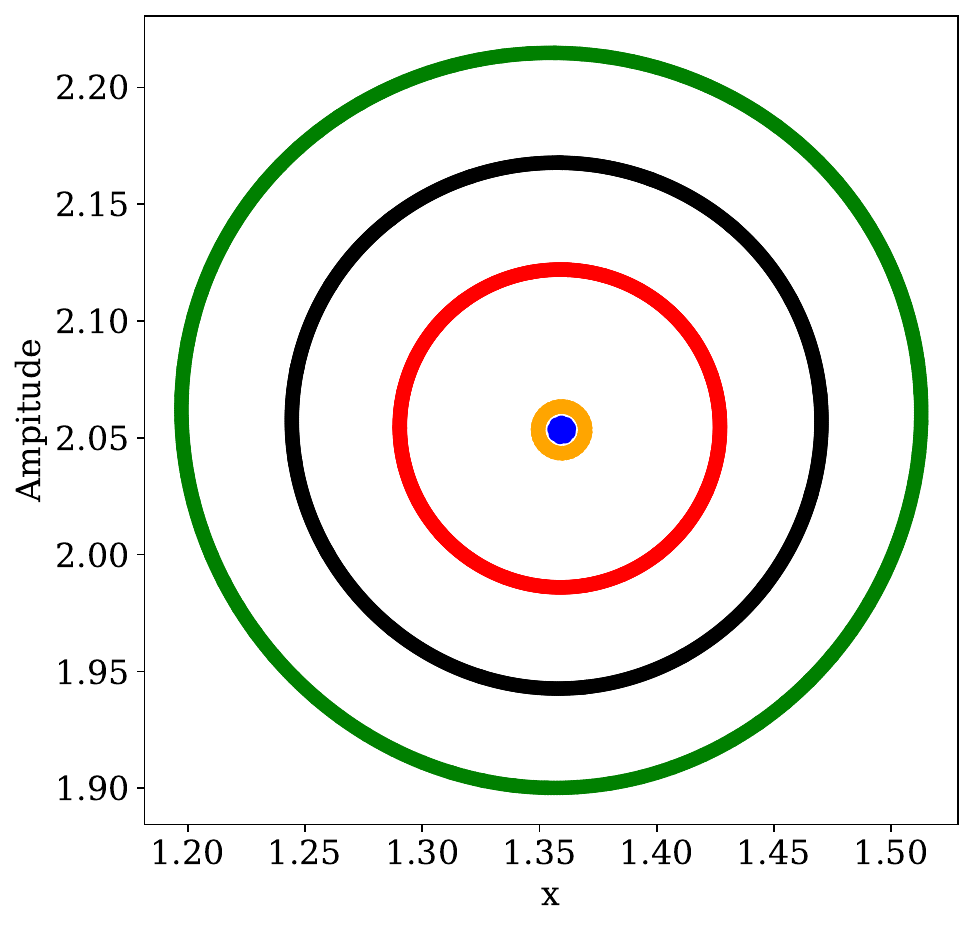}
 	\includegraphics[scale =0.3]{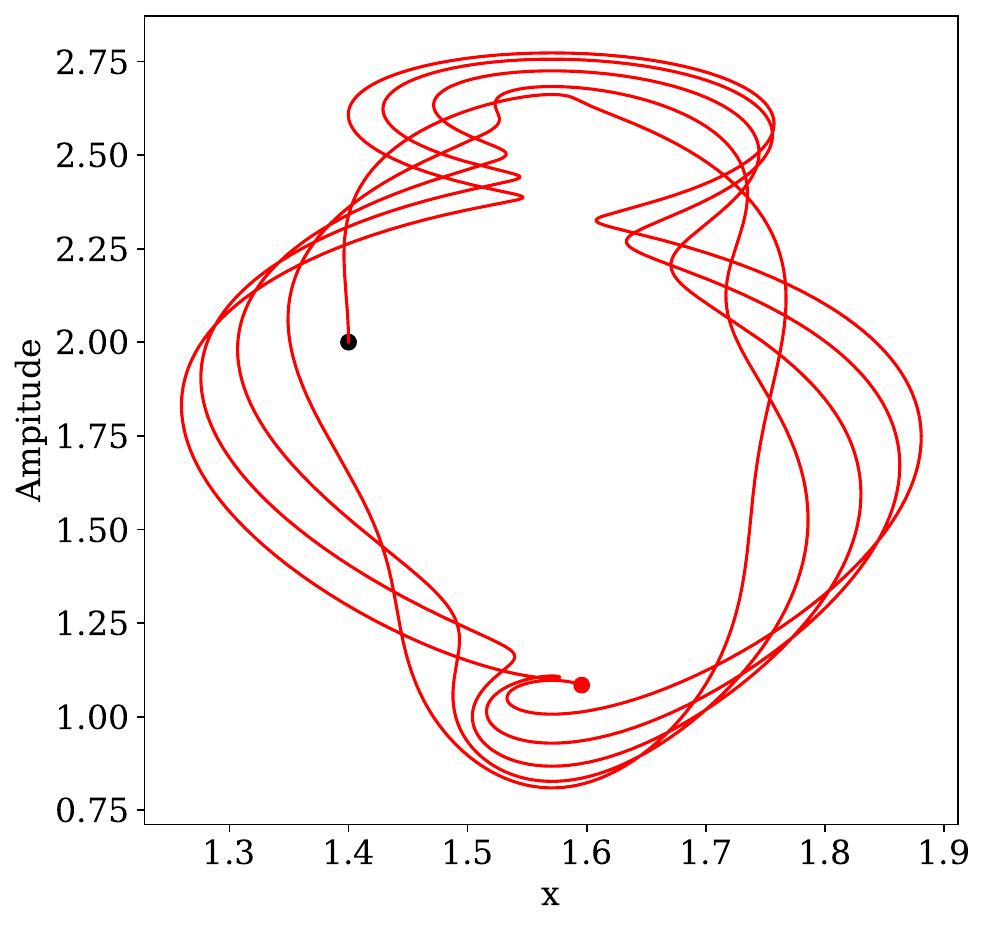}
	\includegraphics[scale =0.3]{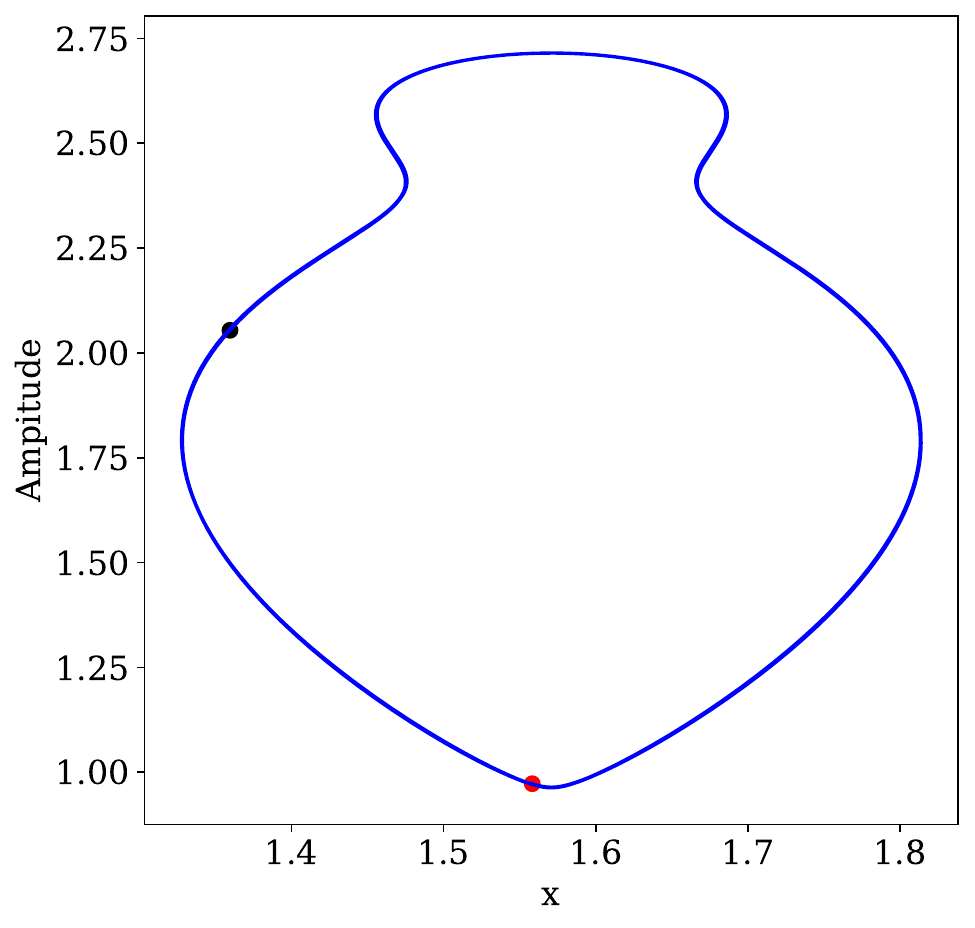}
	\caption{The Poincar{\' e} section (on the left) for five different points and its characteristic solutions are quasi-periodic (in the center) and periodic (on the right). The colors of the trajectory and orbit on the Poincare section match. $A=q=1$, $V_0=-4$, $V=-3$.}.
	\label{Fig001}
\end{figure}

A useful approach for studying this system with an arbitrary set of parameters is the construction of a Poincar{\' e} section (in this case, a stroboscopic map), which can provide insights into the trajectory behavior. This mapping works as follows: it associates the point $(x, a)$ with the point $(\hat{x}, \hat{a})$, which represents the solution of system (\ref{DS_without_broad_force}) at time $t = 2\pi/\omega$ (equal to the period of the change in the external force velocity), given that the solution starts from the point $(x, a, t=0)$.   If the closure of the point's orbit forms a curve on the $(x, a)$ plane, we can describe the soliton motion as quasi-periodic (Fig. \ref{Fig001}, center). If only a single point appears in the Poincar{\' e} section, the solution is periodic (Fig. \ref{Fig001}, right). If neither of these patterns is observed, it is most likely that the soliton exhibits chaotic or transient behavior.

\section{Numerical results}
In contrast to the case with a localized external force, a variable-speed external field induces radiation at a distance from the solitary wave due to interactions between the uniform flow and the changing medium. While this radiation does not impact the leading-order asymptotic analysis, it can complicate numerical simulations. To mitigate this, we utilize the change of variables introduced by Malomed \cite{Malomed:1993}, enabling the decomposition of the wave field into two distinct components as 
\begin{equation}
U(x,t)=u(x,t)+\epsilon u_{0}(x),
\end{equation}
where 
\begin{equation} \label{change}
u_{0}(x)=\frac{A}{\Delta-q^2}\sin(qx),
\end{equation}
is the solution of the linearized forced mKdV equation (\ref{mfKdV}), we have that $u(x,t)$ satisfies 
$$u_{t}+\Delta u_{x}+ 6u^{2}u_x +u_{xxx}=-6\epsilon(u^{2}u_{0})_x+\mathcal{O}(\epsilon^{2}).$$
Consequently, at first approximation we obtain the new equation
\begin{equation}\label{UmfKdV}
u_{t}+\Delta u_{x}+6u^{2}u_x +u_{xxx}=-6\epsilon(u^{2}u_{0})_x,
\end{equation}
where the perturbation now is localized along the free surface $u(x,t)$.  However, it should be noted that when $\Delta$ is time-dependent, the change of variables described in equation (\ref{change}) is no longer applicable. In such cases, equation (\ref{mfKdV}) must be solved directly. This distinction leads to a potential discrepancy between the results of the asymptotic analysis and direct numerical simulations. The asymptotics account for the 'theoretical' soliton amplitude, whereas direct numerical simulations capture the maximum of the wave field, which does not necessarily correspond to the soliton amplitude. Consequently, the observed maximum exhibits 'random fluctuations' as the soliton interacts with and merges into the background field due to the periodicity of the external force.

Equation (\ref{UmfKdV}) and (\ref{mfKdV}) are numerically solved using a standard Fourier pseudospectral method with an integrating factor approach \cite{Trefethen:2000} on a periodic computational domain, $[-L, L]$, with a uniform grid of $N$ points. To minimize the effects of spatial periodicity, such as the reappearance of small waves, we select a domain that is sufficiently large and $\epsilon=0.01$. Thus, the numerical results can be compared to the asymptotic ones. The initial condition for this initial value problem is consistently set as a soliton solution of the mKdV equation (\ref{solitary}) with an amplitude of $a$ and phase $x_0$. In other words,
\begin{equation}\label{solitary}
u(x,0)=a\sech(a(x-x_0)).
\end{equation}
It is important to note that if we select a significantly large soliton amplitude, the effect of the external force on the soliton may become negligible. For this reason, we choose to fix the amplitude at a more moderate level.

\begin{figure}[h!]
	\centering	
	\includegraphics[scale =1.1]{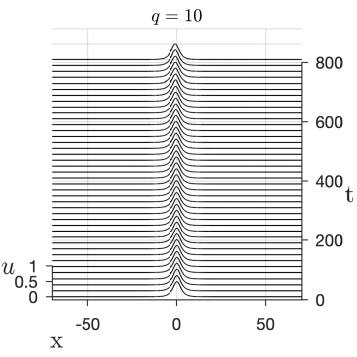}
 	\includegraphics[scale =1.1]{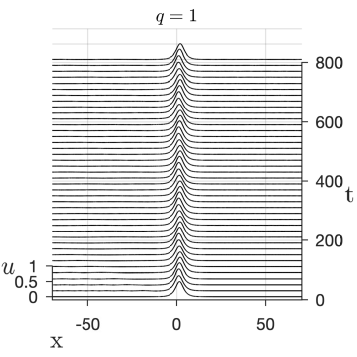}
	\includegraphics[scale =1.1]{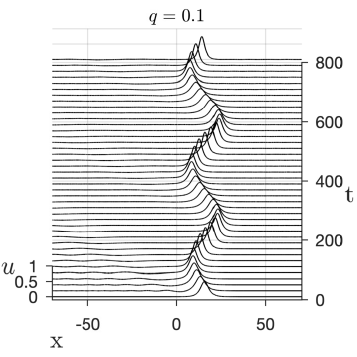}
	\includegraphics[scale =1.1]{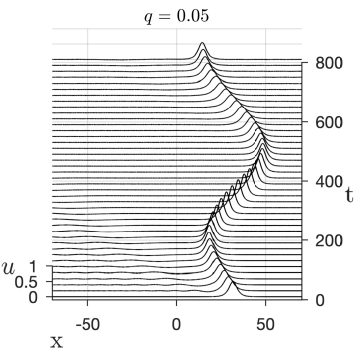}
	\caption{Trapped wave solutions for different values of the wavenumber $q$. The solitary waves are centered at $x_0=\pi/2q$. The deviation speed is $\Delta=-a^2$. Thus, the pair $(x_0,\Delta)$ is a center of the dynamical system (\ref{DS}).}
	\label{Fig3}
\end{figure}

\begin{figure}[h!]
	\centering	
	\includegraphics[scale =1.1]{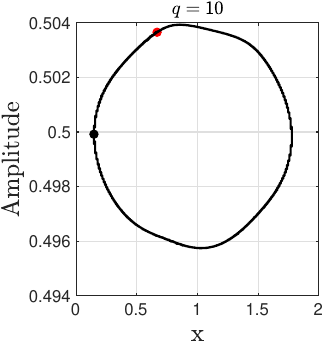}
 	\includegraphics[scale =1.1]{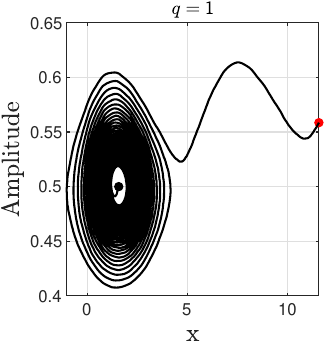}
	\includegraphics[scale =1.1]{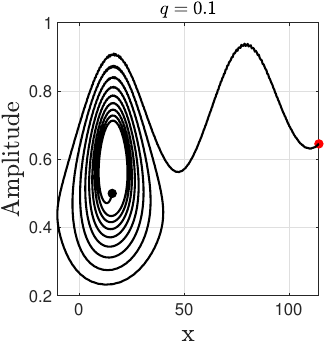}
	\includegraphics[scale =1.1]{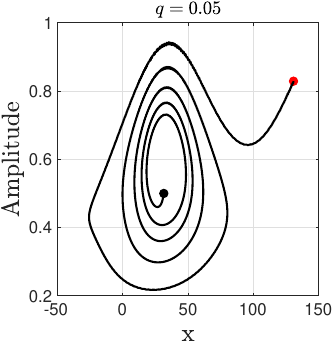}
	\caption{The soliton-amplitude vs. crest position space for the solitons displayed in Figure \ref{Fig3}. The black dot indicates the initial position of the soliton-amplitude and crest and the red one indicates the final position.}
	\label{Fig4}
\end{figure}

We conducted a series of simulations in which the external force was applied with varying frequencies to represent both narrow and broad forcing scenarios. Figure \ref{Fig3} illustrates the results of these numerical simulations for a range of external force frequencies. When the frequency is high, the soliton remains near the central equilibrium point as predicted by asymptotic theory, experiencing only small, localized oscillations around its initial position. However, as the frequency of the external force decreases, the amplitude of the soliton's oscillations around its original position increases, indicating a more pronounced dynamic response. Notably, the soliton exhibits a periodic response to the periodic external forcing, with the soliton crest position over time revealing this behavior in the form of a consistent oscillatory pattern. This frequency-dependent behavior highlights the soliton sensitivity to changes in the external force characteristics, allowing for insights into the soliton's resonant properties and stability under varying forcing conditions.
Another key aspect to analyze is the behavior of the soliton in the soliton-amplitude vs. crest position space. Figure \ref{Fig4} corresponds to the scenarios depicted in Figure \ref{Fig3} and provides a clear visual representation of this behavior. These diagrams reveal that while the soliton initially rebounds near its starting position, it eventually drifts away from the influence of the external force. In this phase space, the soliton's trajectory resembles an unstable spiral, indicating that the soliton oscillates around its initial position with increasing amplitude over time. For high-frequency forcing, however, the soliton exhibits nearly closed orbits, suggesting a more stable periodic behavior. This highlights the soliton's sensitivity to frequency, where large frequencies promote bounded oscillations, while lower frequencies lead to unbounded, progressively divergent motion.

\begin{figure}[h!]
	\centering	
	\includegraphics[scale =1]{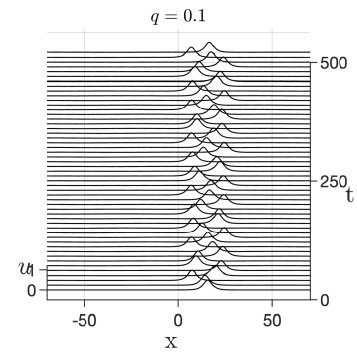}
		\includegraphics[scale =1]{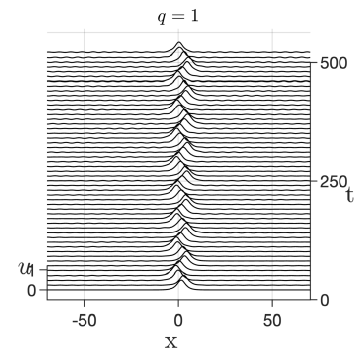}
	\caption{Left: The soliton-amplitude vs. crest position space for the solitons for the resonant frequency. The dot indicates the final position of the soliton-amplitude and crest. Right: The soliton amplitude as a function of time. Here, $x_0=\pi/2q$, $V_0=a_0^2$, $V=-1.5$ and $\omega=\omega_0$.}
	\label{Figres1}
\end{figure}

\begin{figure}[h!]
	\centering	
		\includegraphics[scale =1.1]{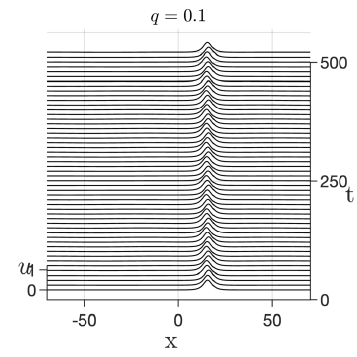}
		\includegraphics[scale =1.1]{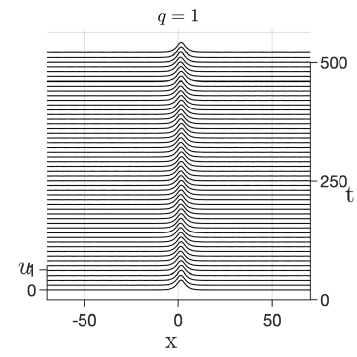}
	\caption{Left: The soliton-amplitude vs. crest position space for the solitons for the resonant frequency. The dot indicates the final position of the soliton-amplitude and crest. Right: The soliton amplitude as a function of time. Here, $x_0=\pi/2q$, $V_0=a_0^2$, $V=0.1$ and $\omega=\omega_0$.}
	\label{Figres2}
\end{figure}

Finally, we investigate soliton dynamics when the speed of the external force oscillates at the resonance frequency, as determined in the linear approximation (see equation \ref{resonance})). Since our system is nonlinear, the soliton amplitude does not increase linearly over time as it would in a purely linear system.{ Figures \ref{Figres1} and \ref{Figres2} illustrate the soliton evolution at the resonant frequency for different values of $q$. As observed, the soliton behavior closely resembles the case of constant 
$\Delta$, particularly for small values of $V$ (see Figure \ref{Figres2}). Although the background field is small compared to the soliton amplitude, it still influences the soliton by introducing minor fluctuations that cannot be mitigated using the change of variables previously mentioned. Nevertheless, the asymptotic and numerical results show qualitative agreement at early times. }

\section{Conclusion}
In this work, we have investigated the interactions of solitons with a periodic external field within the modified Korteweg-de Vries (mKdV) framework, employing both asymptotic analysis and direct numerical simulations. Interestingly, we found that the asymptotic results qualitatively diverge from the behavior observed in the numerical simulations. Specifically, while the asymptotic approach predicts an equilibrium center and closed orbits in the soliton-crest vs. soliton-amplitude phase space, the numerical simulations reveal unstable spiral trajectories under the same conditions. Notably, however, the soliton remains near its initial position during early times, consistent with both approaches. 

To enhance the accuracy of the asymptotic results, two potential modifications could be explored. First, extending the asymptotic expansion to include higher-order terms may improve the approximation. Second, introducing a viscosity term into the model would fundamentally alter the dynamics, potentially transforming the equilibrium center points into stable or unstable spirals, depending on the sign and magnitude of the viscosity. These adjustments represent promising directions for future research, as they could provide deeper insights into the soliton behavior under periodic forcing and bring the asymptotic and numerical results into closer agreement.

\section{Acknowledgements}
The article was prepared within the framework of the project <<International academic cooperation>> HSE University.

	\section*{Declarations}
	
	\subsection*{Conflict of interest}
	The authors state that there is no conflict of interest. 
	\subsection*{Data availability}
	
	Data sharing is not applicable to this article as all parameters used in the numerical experiments are informed in this paper.

\end{document}